\documentclass[aps,twocolumn,nofootinbib]{revtex4}

\usepackage{graphicx}
\usepackage{amssymb, amsmath,mathtools}
\usepackage{bbm}
\usepackage{enumerate}

\graphicspath{{./}{./plots/}}

\usepackage{xcolor}

\begin{document}

\title{Chiral symmetry breaking in three-dimensional quantum electrodynamics as fixed point annihilation}

\author{Igor F. Herbut}

\affiliation{ Department of Physics, Simon Fraser University, Burnaby, British Columbia, Canada V5A 1S6}

\begin{abstract}
Spontaneous chiral symmetry breaking in three dimensional ($d=3$) quantum electrodynamics is understood as annihilation of an infrared-stable fixed point that describes the large-N conformal phase by another unstable fixed point at a critical number of fermions $N=N_c$. We discuss the root of universality of $N_c$ in this picture, together with some features of the phase boundary in the $(d,N)$ plane. In particular, it is shown that as $d\rightarrow 4$, $N_c\rightarrow 0$ with a constant slope, our best estimate of which suggests that $N_c = 2.89$ in $d=3$.

\end{abstract}

\maketitle

Three-dimensional (non-compact) quantum electrodynamics ($QED_3$) has served as a paradigmatic example of a strongly coupled relativistic gauge field theory \cite{pisarski}, and has been both of methodological importance in high-energy physics, and with applications in modern condensed matter physics. It is generally accepted that at a large number ($N$) of Dirac fermions the theory features a conformal interacting phase characterized by  a set of non-trivial power-law correlations \cite{strack, klebanov}. At small $N$, on the other hand, chiral symmetry is believed to be spontaneously broken, and consequently, the fermionic spectrum is believed to be gapped \cite{appelquist1, nash}.   The critical value of $N$ that separates the conformal from the chiral symmetry broken (CSB) phase has been a matter of three-decades-long debate, with the analytically computed numbers ranging anywhere from one to ten \cite{appelquist2, raya, maris, fischer, braun, raviv, pietro, chester, janssen, kotikov}.  Numerical lattice calculations have been utilized from the very beginning  \cite{dagotto, hands}, with the most recent one \cite{narayanan} adding to the controversy by not finding the condensate even when $N=1$. Different versions of the higher-dimensional c-theorem have also been brought to bear \cite{appelquist3, grover}, with the latest bound \cite{giombi} being the tightest: $N_c < 4.4$. The matter is not only of mathematical interest, since in a majority of its applications to strongly correlated electronic matter the number of four-component Dirac fermions is fixed at $N=2$, \cite{rantner, franz, herbut, hermele} and therefore how precisely the value of $N_c$ stands in relation to this number directly influences the phase diagram of some of the most intriguing low-dimensional condensed matter.

Analytical approaches have often been confined to solving the coupled Schwinger-Dyson equations in space-time dimension $d=3$ in progressively more sophisticated approximations. Here we entertain an alternative picture, in which the instability of the conformal phase is understood as a collision between an infrared fixed point that represents it in some space of coupling constants, with another, critical fixed point, which approaches it as $N\rightarrow N_c + $. Indeed, the characteristic Miransky-Kosterlitz-Thouless dependence of the fermion mass on the fermion number found in many large-N calculations \cite{appelquist1}  is strongly suggestive of this ``fixed-point collision and annihilation" scenario \cite{kubota, kaveh, jaeckel, kaplan}, in which the two fixed points after the collision would typically become complex, and that way leave behind only a runaway flow in the physical space of real coupling constants.  The same mechanism has been used in the studies of the phase diagram of non-Abelian gauge theories \cite{jaeckel, kaplan, jbraun}, and was proposed recently for a nematic instability of the gapless semiconductors with nonrelativistic Coulomb interactions \cite{hj, jh}. An attractive feature of this description is that the universality of the critical number $N_c$ in $d=3$ derives directly from the standard universality of the correlation length critical exponent $\nu$ at the second fixed point: assuming that the colliding fixed point, in a general dimension $d$ and at fermion number $N$, is the standard critical point with a single infrared-relevant direction, yields an equation for the location of the  phase boundary between the conformal and the CSB phases:
\begin{equation}
\nu ^{-1} (d,N) =0.
\end{equation}
Although the fixed point values of the coupling constants in general are not universal, the eigenvalues of the stability matrix at the fixed points, such as the inverse of the correlation length critical exponent $\nu$, are, and this at the point of collision and in $d=3$ translates into the critical number of fermions $N_c (3)$.

The potential profitability of this approach to the problem becomes apparent when one realizes that, quite generally, as $d\rightarrow 2$ the critical number $N_c \rightarrow \infty$ \cite{janssen, hj}.  This is because {\it any} contact interaction between the fermions at a noninteracting fixed point has a dimension of $2-d$.  The exactness of the one-loop calculation at $N=\infty$, and the fact that the charged fixed point is then at $e^2 _* \sim 1/N$  dictate that at any critical point in this limit $\nu =  1/(d-2)$.  The second useful limit is $d\rightarrow 4$, in which, as it will be shown here,  $N_c \rightarrow 0$ with a constant slope. The value of this slope should provide an independent useful estimate of the value of $N_c$ in $d=3$, and will be the main subject of our calculation. We argue that although the slope of the phase boundary near the point $(d=4, N=0)$ in principle depends on all orders in perturbation theory in powers of gauge coupling, a reasonable result may be obtained already at the second-order approximation.

To be specific, we focus on dimensions close to $d=4$ and consider the following field theory for massless Dirac fermions with four-fermi interactions, and coupled to the $U(1)$ gauge field:
\begin{equation}
L= \bar{\Psi}_n i \gamma_\mu  ( \partial_\mu - i e A_\mu) \Psi_n + \sum_{a=1}^2 g_a (\bar { \Psi}_n X_a \gamma_\mu \Psi_n)^2 + \frac{ F_{\mu\nu}^2}{4}
\end{equation}
where $\mu,\nu = 0,1,2,3$, $n=1,...N$,  the summation over the Greek indices is assumed, and $X_1= 1$ and $X_2 = \gamma_5 = \gamma_0 \gamma_1 \gamma_2 \gamma_3$. The Lagrangian is invariant under the local $U(1)$ gauge symmetry, but also under the global $[U(1)] ^N = U(1)  \times U(1)\times.... U(1)$ chiral transformation,  $\Psi_n \rightarrow exp (i\theta_n \gamma_5) \Psi_n$. We assume Euclidean metric, natural units with $\hbar = c =1$, and a Hermitian irreducible (four-dimensional) representation of the Clifford algebra $\{ \gamma_\mu, \gamma_\nu \} = 2 \delta_{\mu\nu}$. The calculations will, for convenience, be performed in Feynman gauge, but the results should be independent of this particular choice.

With the decrease of the UV cutoff $ \Lambda \rightarrow \Lambda/b $ the three coupling constants flow according to the one-loop equations:
\begin{equation}
\beta_1 = (2-d) g_1 + 4 (N+1) g_1 ^2 - 8 g_1 g_2 - 6 e^2 g_2,
\end{equation}
 \begin{equation}
\beta_2 = (2-d) g_2 + 2 (2N-1) g_2 ^2 + 4 g_1 g_2 - 6g_1 ^2 - 6 e^2 g_1 - \frac{3}{2} e^4,
\end{equation}
\begin{equation}
\beta_e = (4-d) e^2 + \beta_{e0} (e) .
\end{equation}
Here $\beta_a = dg_a/d\ln(b)$, $\beta_e= de^2 / d\ln (b)$, with $\beta_{e0} (e) $ as the latter in dimension $d=4$. The beta-function $\beta_{e0}(e) $ in $d=4$ is known to four-loop \cite{gorishny}, and even five-loop order \cite{kataev},
\begin{equation}
\beta_{e0} (e) = -\frac{4N}{3} e^4 - 4 N e^6 + O(N e^8, N^2 e^8),
\end{equation}
but we display only the leading two terms in the expansion, which are also the only ones with universal coefficients. In the above we have rescaled all the couplings as $g_a S_d \Lambda^{d-2} /(2\pi)^d \rightarrow g_a$, and $e^2 S_d \Lambda^{d-4} /(2\pi)^d \rightarrow e^2 $, with $S_d = 2\pi^{d/2} / \Gamma (d/2)$ as the area of the unit sphere in $d$ dimensions.

The crucial feature of the above equations is the last term in Eq. (4), which is independent of both four-fermi couplings $g_a$. Its presence implies that the coupling $g_2$ is {\it generated}  when $e^2 \neq 0$ by the change of the cutoff, even if absent initially. Once $g_2$ becomes generated, the last term in Eq. (3) generates $g_1$ as well. No other contact interactions are then further introduced in the process of mode elimination, and the theory in Eq. (2) is closed under the renormalization group (RG). This requirement in fact dictates the specific choice of the four-fermi interactions in the Lagrangian. The two four-fermi terms in Eq. (2) may also be understood as being precisely the four-dimensional equivalents of the corresponding four-fermi terms in three dimensions \cite{pietro}.

Let us first observe two general features of the flow equations when $e^2=0$ implied by the symmetry. First, in this limit the above equations feature $g_1=g_2$ as an invariant line under the RG. At this line the upper ($u_n$) and the lower ($v_n$) spinors of four-component fermion fields $\Psi_n ^T = (u_n,v_n)$ decouple: in the representation in which the three matrices $\alpha_k = i\gamma_0\gamma_k = \sigma_3 \otimes \sigma_k$, and the quadratic part of the action becomes block diagonal, the remaining two matrices are $\gamma_0=\sigma_1 \otimes 1$, and $\gamma_5 =\sigma_3 \otimes 1$. $\sigma_k$ are the standard Pauli matrices.
It is easy to check that  for $g_1=g_2$ the four-fermi interaction term becomes:
\begin{eqnarray}
\sum_{a=1}^2 g_a (\bar { \Psi}_n X_a \gamma_\mu \Psi_n)^2=
2 g_1 ( ( u^\dagger _n u_n)^2 + \\ \nonumber
(v^\dagger _n v_n)^2 -  ( u^\dagger _n \sigma_i u_n)^2 -(v^\dagger _n \sigma_i v_n)^2) ,
\end{eqnarray}
and thus the upper and lower spinors are in this limit fully decoupled at {\it any} cutoff. The line $g_1=g_2$, $e^2=0$ is therefore not only one loop, but an exact RG invariant.

The second exact RG invariant is $e^2=0$, $g_1=0$, as also suggested by the one-loop equations. This is due to a hidden enhanced symmetry in this limit. Assume a (different) representation in which the three matrices $\alpha_k = i\gamma_0 \gamma_k$, $k=1,2,3$,  featured in the Dirac Hamiltonian are all real. The matrices $\gamma_0$ and $\gamma_5 = i \alpha_1 \alpha_2 \alpha_3$ are then both imaginary, and Hermitian. \cite{clifford} If one forms the usual Nambu-Gor'kov doublet as $\Phi_n (p) ^T = (\Psi_n (p), \Psi_n ^* (-p) )$, the Lagrangian at $g_1=e^2=0$ may be rewritten as
\begin{widetext}
\begin{equation}
L= \frac{1}{2} \Phi_n ^\dagger ( \partial_\tau + (1\otimes \alpha_k) (-i\partial_k) ) \Phi_n +
\frac{g_2}{4} [ \sum_{k<j} ( \Phi_n ^\dagger (1\otimes \alpha_k \alpha_j)\Phi_n )^2 -  (\Phi_n ^\dagger (1\otimes \alpha_1 \alpha_2 \alpha_3 )\Phi_n ) ^2].
\end{equation}
\end{widetext}
The block-diagonal form in this representation implies that the symmetry in this limit becomes $ [ U(2) ]^N$, with each $U(2)$ factor generated by
$( 1\otimes i \alpha_1 \alpha_2 \alpha_3,  \sigma_3 \otimes 1, \sigma_1 \otimes i \alpha_1 \alpha_2 \alpha_3 , \sigma_2 \otimes i \alpha_1 \alpha_2 \alpha_3 ) $. The reader can easily check that when $g_1 \neq 0$, or $e^2\neq 0$,  only the first two of the four generators remain unbroken, and each factor of $U(2)$ is reduced to $U(1)\times U(1)$. These are then the original chiral and particle-number symmetries, the latter being left implicit in Eq. (2).

 Eqs. (3)-(5) feature several fixed points for general $N$ and $d$. The results become particularly transparent when $4-d \ll 1$. It is more convenient to consider the combinations $g_\pm = g_1 \pm g_2$, and rewrite the RG equations as
 \begin{equation}
 \beta_+ = (2-d) g_+ + 2 (N-1) g_+ ^2 + 2N g_- ^2 -6 g_+ e^2 -\frac{3}{2} e^4,
 \end{equation}
 \begin{equation}
 \beta_- = (2-d) g_- + 6 g_- ^2 +  4(N+1) g_+ g_-  + 6 g_-  e^2 + \frac{3}{2} e^4.
 \end{equation}
 If we focus on the limit $N\rightarrow 0$, which will be relevant when $d\rightarrow 4-$, the first equation conveniently decouples. In this limit, for $e^2=0$ there are four fixed points: 1) stable, Gaussian, at $g_\pm =0$, 2) critical, at $g_+=0$, $g_- = 1/3$, 3) critical, at $g_+ = -1$, and $g_- =0$, and 4) bicritical, at $-g_+ = g_- = 1$. The last two fixed points lie on the high-symmetry invariant lines identified above, and will be unimportant for the following analysis. Turning on a small charge $0<e^2\ll 1$ then shifts the fixed points 1) and 2) toward each other, while 3) and 4) move away from the origin. The shifted, stable, Gaussian fixed point may be understood as representing the conformal phase of QED, and the fixed point 2) may be understood as being the critical point for  spontaneous CSB,  that would be induced by increasing the coupling $g_-$ above a (positive) threshold. This interpretation becomes particularly appealing upon realizing that for $N=1$, for example, when $g_+=0$ the quartic interaction term in the theory by Fierz identity becomes:
 \begin{equation}
 \sum_{a=1}^2 g_a (\bar { \Psi} X_a \gamma_\mu \Psi )^2 = - g_- [ (\bar{\Psi}\Psi )^2 - ( \bar{\Psi} \gamma_5 \Psi)^2 ] ,
 \end{equation}
so that a large positive coupling $g_- $ clearly energetically favors the formation of finite chiral condensate, i. e.  having
$\langle \bar{\Psi} e^{i \theta \gamma_5} \Psi \rangle \neq 0$.

 Furthermore, when $N \rightarrow 0$, the function $\beta_+$ for any value of $e^2$ has two real, negative zeros in $g_+$. The function $\beta_-$, on the other hand, allows real zeros for the second coupling $g_-$, in the same limit $N\rightarrow 0$,   only if
 \begin{equation}
 1 - 2 g_+ - 6 e^2 > 0.
 \end{equation}
 Using the solution of $\beta_+ =0$ we then find the condition for the critical value of the charge at which the two zeros of $\beta_-$, corresponding to  the shifted fixed points 1) and 2), turn complex:
\begin{equation}
 e_c ^4 - 6 e_c ^2 + 1=0,
 \end{equation}
with the smaller of the two roots determining $e_c^2 = 3- 2\sqrt{2}= 0.17157$. The critical value of $N$ in this limit may now finally be extracted from Eq. (5) by equating the critical value for the collision of the two fixed points with the infrared-stable fixed point value of $e^2$ at which $\beta_e =0$. This yields,
\begin{equation}
\frac{4-d}{N_c} = - \lim_{N\rightarrow 0} \frac{ \beta_{e0} ( e_c )} {N e_c^2}.
\end{equation}
Note that the limit on the right-hand side eliminates the terms in the function $\beta_{e0} (e)$ that are nonlinear in $N$, which are not displayed in Eq. (5), but which do appear beginning at the third loop.

We use the right hand side of the last equation to the second order in $e^2 _c$, to finally obtain the critical number of fermion components in the vicinity of $d=4$:
\begin{equation}
N_c  = \frac{3 (4-d) } { 4 (e_c^2 + 3 e_c ^4) } \approx  2.88596 (4-d) +  O ( (4-d)^2 ).
\end{equation}

Several comments on the nature of the above procedure are in order. First, we note that the CSB transition in the theory at $e^2=0$  with only four-fermi interaction has the upper critical dimension of $d=4$, so terminating the expansions in $g_a$ at one loop when $d\rightarrow 4$ and at $e^2 =0$ is entirely justified, and yields the correct (mean field) values of all critical exponents. Moreover, by increasing the value of the charge by hand we find the collision of the stable fixed point and the CSB critical point occurring at $g_+ = - e_c^4 / 2= -  0.0147 $, and $g_- =e_c^2 /2 = 0.0857$, safely in the weak coupling region, so that the terms containing the four-fermi couplings $g_\pm $ in $\beta_{e0}(e)$, which would be of order $O(g_\pm ^2)$, should indeed be negligible. Everything hinges, however, on the numerical smallness of the critical value of the charge $e_c^2$, which, although reasonably well fulfilled by the present computation, is by no means guaranteed. It would be more systematic  to regard the solution for the critical charge in the limit of $N\rightarrow 0$ and $d\rightarrow 4$ as actually being
\begin{equation}
e_c ^2 = \frac{1}{6} + O(e_c ^4),
\end{equation}
where only the leading term of 1/6 is unambiguously determined by the one-loop RG equations (9)-(10).  Using $e_c ^2 =1/6$ and Eq. (14) with the two-loop result for $\beta_{e0}$ yields then $N_c = 3 (4-d)$. Staying faithful  to the first-order  perturbation theory and ignoring the sizable $e_c^4$ term in $\beta_{e0}(e) $ as well, would give a higher value of $N_c = (9/2) (4-d)$, slightly above the bound in \cite{giombi}.  This discussion should make it clear that the slope of the $N_c(d)$ line near $d=4$, strictly speaking, is not a simple perturbative quantity. It may turn out to be effectively perturbative, however, by a fortunate accident that the critical value $e_c ^2$ for the fixed point collision happens to be reasonably small. This is the case in the actual one-loop calculation, and we suspect that the result in Eq. (14) may yield a useful estimate of $N_c$ in the physical dimension of $d=3$.

A related, albeit not quite the same, strategy to calculating $N_c$ was recently employed in ref. \cite{pietro}. The authors' result  may be recovered within the present calculation by neglecting $\sim e^4$ terms in Eqs. (9) and (10), and considering the Gaussian and CSB critical points, which are then both in the $N\rightarrow 0$ limit always on the line $g_+ =0$. Since the Gaussian fixed point then remains pinned at $g_- =0$, the critical point goes through it when
$1-3 e_c ^2=0$. Using the leading-order result for $\beta_{e0}$ then yields $N_c = (9/4) (4-d)$ when $d\rightarrow 4$ \cite{pietro}. This way the CSB critical point would not become complex for $N<N_c$, but would only move to the region $g_- <0$, destabilizing en passant the Gaussian fixed point.

 When the fixed point values of the couplings are complex but still with small imaginary parts, the RG trajectories linger around the region where the annihilation took place. The long RG time it takes to escape this region translates into a small fermion mass near the $N_c(d)$ phase boundary. Solving the RG equations in the simplest approximation in which $g_+$ is completely neglected and the gauge coupling set at its fixed point value, for example, yields the fermion mass $m$ close to the phase boundary to be
 \begin{equation}
 \ln \frac{\Lambda}{m} = \frac{2 \pi}{ (d-2) \sqrt{1 -(N/N_c (d)) } } + O(1),
 \end{equation}
 with $N_c (d) = 9 (4-d)/(d-2)$ in this approximation. This is the characteristic Miransky-Kosterlitz-Thouless essential singularity in the control parameter, which is a universal characteristic of the ``fixed point collision and annihilation" scenario \cite{kaplan}. The numerical value of the numerator of the right-hand side is nevertheless nonuniversal \cite{correction}.

In conclusion, we hope to have shown how the instability of the conformal phase of the $QED_d$ in dimensions $2<d<4$ towards chiral symmetry breaking can also be phrased as an annihilation of a  conformal stable fixed point by a critical fixed point, in the space of two specific four-fermi couplings. The difficulty in pinning down the value of the critical number of fermions $N_c$ in $d=3$ is related to the fact that as $d$ is changed from four to two, $N_c(d)$  varies, presumably smoothly, between zero and infinity. We have shown that the slope of the phase boundary $N_c(d)$ when $d\rightarrow 4$ is finite, and estimated its value using the  two-loop polarization in $QED_4$. The extrapolation of this straight line from $d=4$ to $d=3$ suggests $N_c(3) =2.89$.

The author is grateful to Igor Boettcher, Lukas Janssen, Andrei Kataev, Igor Klebanov, Adam Nahum, Michael Scherer, and Philipp Strack, for useful discussions and correspondence. This work was supported by the NSERC of Canada.

\end{document}